\documentclass[preprint]{aastex}
\usepackage{graphicx}
\usepackage{subfigure}
\usepackage {natbib}

\begin{document}

\title {WATER FRACTIONS IN EXTRASOLAR PLANETESIMALS}

\author{M. Jura\altaffilmark{a} \& S. Xu(\includegraphics[width=1.2cm]{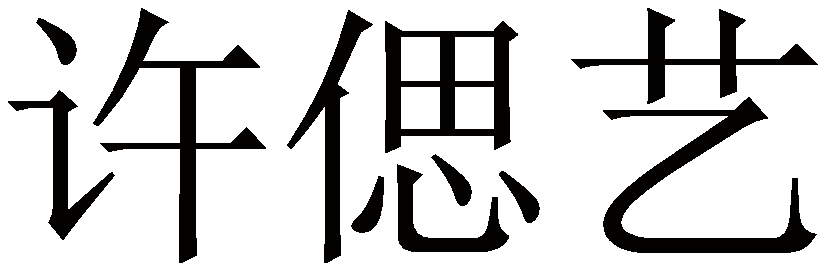})\altaffilmark{a}} 

\altaffiltext{a}{Department of Physics and Astronomy, University of California, Los Angeles CA 90095-1562; jura, sxu@astro.ucla.edu}

\begin{abstract}

With the goal of using externally-polluted white dwarfs to investigate the water fractions of extrasolar planetesimals,  we assemble from the literature a sample  that we estimate to be more than 60\% complete of  DB white dwarfs warmer than 13,000 K, more luminous than 3 ${\times}$ 10$^{-3}$ L$_{\odot}$ and  within 80 pc of the Sun.    When considering all the stars together, we find the summed  mass accretion rate of heavy atoms  exceeds that of hydrogen by  over a factor of 1000.   If so, this sub-population of extrasolar asteroids treated as an ensemble has little water and is at least a  factor of 20  drier than  CI chondrites, the most primitive meteorites.  In contrast, while an apparent ``excess" of oxygen in a single DB can be interpreted as evidence that the accreted material originated in  a water-rich parent body, we show that at least in some cases, there can be sufficient uncertainties in the time history of the accretion rate that such an argument may be ambiguous.  Regardless of the difficulty associated with  interpreting the results from an individual object,  
our analysis of the population of polluted DBs provides indirect observational support for the theoretical view that a snow line is important in disks where rocky planetesimals form.
 \end{abstract}
\section{INTRODUCTION}
Evidence is now strong that most externally-polluted white dwarfs  
derive their heavy atoms\footnote{We use ``heavy atoms"  to mean all elements with atomic number greater than two.  We do not follow astronomical convention and use ``metals" for this matter because of this word's  different meaning in the context of the physics of planets.} by accretion from disrupted planetesimals \citep{Jura2003, Gaensicke2006, Jura2008, Farihi2010, Zuckerman2010}.  Observational progress is rapid; both detailed studies of individual
objects \citep{Klein2010,Klein2011,Dufour2010,Farihi2011a, Farihi2011b,Vennes2010,Melis2011,  Zuckerman2011} and broader  surveys  \citep{Koester2011,Kawka2011b,Zuckerman2010} are now being performed.  Here, we combine recent studies of white dwarfs to assess the water fraction  -- either in pure form or in hydrated minerals \citep{Rivkin2002} -- within an ensemble of extrasolar minor planets.  

Oxygen is so cosmically abundant that   H$_{2}$O is predicted to  be a major constituent in solid planets  that form in regions   where the temperature in the planet-forming disk  is sufficiently low that ice is stable.   Water is widespread in the outer
solar system \citep{Jewitt2007,Encrenaz2008}, and can be more than 50\% of the mass of Kuiper Belt Objects and comets.  Ceres, the largest asteroid, is probably  ${\sim}$25\% water \citep{McCord2005,Thomas2005} and CI chondrites are nearly 20\% water \citep{Wasson1988}.  In contrast, much of the inner solar system is dry.   For example, the refractory-rich CV chondrites   are only about 2\% water \citep{Wasson1988}.    Although the amount  of   internal water is not well known, the water fraction of bulk Earth is between 0.06\% and  2\%  \citep{vanThienen2007}. The usual theoretical explanation for the large radial variation in water content in the solar system is that there is a snow line in the  planet-forming disk and that ice forms and condenses into planetesimals only in the cold, outer regions \citep{Encrenaz2008}.  Water may be common in some extrasolar planets \citep{Fortney2007,Seager2007,Sotin2007,Valencia2007},  but actual evidence of ice-rich  extrasolar planets is sparse and uncertain \citep{Charbonneau2009,Gould2010,Tinney2011}.

 Because most internal water can survive within an asteroid  with a radius greater than 60 km  during a 3 M$_{\odot}$ star's pre-white-dwarf Asymptotic Giant Branch  (AGB) evolution \citep{Jura2010},
measurements of  oxygen in an externally-polluted white dwarf's atmosphere\footnote{\citet{Gaensicke2010} reported dredged-up oxygen in two massive, cool white dwarfs, but there is no known way that the systems we discuss here have internally-polluted atmospheres.} can serve as a tool to assess the amount of water in an accreted  parent body.
\citet{Klein2010} showed that the oxygen accreted by GD 40 could have been carried in oxides such as MgO and SiO$_{2}$ and that water is less than 10\% by mass of the photospheric pollution.  
 Similarly, in HS 2253+8023 less than 10\% of the accreted mass is water \citep{Klein2011}.    However, because of the inherent uncertainties in abundance determinations, there is little prospect of using this
  technique to reduce much below ${\sim}$10\% by mass the  bound on the amount of water carried in the parent bodies.   
    
  Water-rich parent bodies might be identified if there is
  an excess of oxygen \citep{Jura2010}.  \citet{Farihi2011a} proposed that GD 61 has accreted a water-rich asteroid, but, as discussed in Section 3 below, we argue that  the evidence in support of this suggestion
  is inconclusive because their argument is somewhat sensitive to the unknown time history of the accretion rate.

Another pathway to study the water content in extrasolar asteroids is to measure the amount of hydrogen that results from planetesimal accretion onto DB
white dwarfs\footnote{DBs are white dwarfs where the helium lines dominate the spectrum.  Related objects are the DBAs where hydrogen also is detected and the DZs which are thought to have helium-dominated atmospheres but are too cool for helium lines to appear in the spectrum.} because this light gas never gravitationally settles \citep{Koester2009}.  
At least 1/3 of DBs warmer than 14,000 K are externally-polluted
 by heavy atoms derived from ancient planetesimals \citep{Zuckerman2010}.  These same stars also acquire whatever hydrogen that was bound into water in the parent body.  By assessing the relative amounts of hydrogen
 and heavy-atoms in the stellar photospheres, we can constrain the amount of water in the accreted planetesimals.   For example, \citet{Zuckerman2010} showed that the upper limit to the hydrogen abundance in G241-6 indicates that little water has been accreted.
 
 Because hydrogen  accumulates over the entire cooling age of the star and because the white dwarf may have substantial primordial hydrogen \citep{Bergeron2011}, we cannot attribute the observed hydrogen to any particular accretion event.  Instead, we consider a well-defined ensemble of
 white dwarfs and sum the individual hydrogen accumulation rates to compare with the summed heavy atom  accretion rates.  While particular white dwarfs such as GD 362 have acquired 
 so much hydrogen that they may have accreted ice-rich matter \citep{Jura2009}, here we consider the warm DBs in aggregate.  By considering the ensemble as a whole, this procedure can probe much smaller water fractions than allowed by assessing the amount of hydrogen in individual objects.  Below, we show that water is less than 1\% of the aggregate mass in our sample of extrasolar asteroids, an order of magnitude improvement over what can be achieved by examining
 each star on a case-by-case basis.

  In Section 2, we describe a volume-limited sample of
 DBs and compare the summed accretion rates of both heavy atoms and hydrogen.  In aggregate, the accreted material is dry.  In Section 3, we  describe models for a quasi-steady state accretion and  argue that the evidence that GD 61 is accreting ice-rich material is inconclusive.   In Section 4 we discuss some further
 implications of our analysis, and in Section 5 we present our conclusions.

 \section{AGGREGATE  ACCRETION RATES}
 
 \subsection{The Sample of Stars}
 
 In order to assess the relative amounts of accreted hydrogen and heavy elements, we assemble a volume-limited sample of DBs that is as bias-free as possible.  
 We cannot use the available survey of white dwarfs within 20 pc of the Sun that is nearly complete because it contains  only one DB  \citep{Holberg2008}.  As a compromise
 between having enough stars to perform an analysis yet not being too incomplete, we adopt the outer boundary of our sample at 80 pc. Because it is
 difficult to distinguish helium in atmospheres cooler than 13,000 K and such cool DBs are difficult to characterize; we only consider stars warmer than this temperature.  
 
 Using available literature, we list in Table 1 57 DBs  with luminosities greater than 3 ${\times}$ 10$^{-3}$ L$_{\odot}$ or $M_{Bol}$ = 11 mag, along with  their distances, temperatures, galactic latitudes, masses, and accretion rates.  
 According to \citet{Bergeron2011}, the space density of such DBs
   is 5.15 ${\times}$ 10$^{-5}$ stars pc$^{-3}$,  and we might expect to identify 110 appropriate objects within 80 pc.   We now assess the incompleteness of our sample in  more detail.
   
 If  the stars are distributed isotropically in the sky, we would expect 50\% to  lie at $|b|$ ${\leq}$ 30$^{\circ}$.  However, we only identify 17 stars in this zone.   The incompleteness of Table 1 is at least partly  because
 some stars near the Galactic Plane are missed. As demonstrated by \citet{Lepine2011} and as illustrated by the recent discovery of a  DAZ at a distance of 55 pc with $b$ = -8$^{\circ}$ and $T_{eff}$ = 20,900 K \citep{Vennes2010}, there are nearby warm white dwarfs yet to be found in this region.  
 
In actuality, we expect  fewer than half the stars lie at $|b|$ $>$ 30$^{\circ}$ because the space
density of stars decreases away from the Galactic Plane with a scale height, $h$, which depends upon the mass of the main-seqeunce progenitor.  Assume that
 \begin{equation}
n\;=\;n_{0}\;e^{-|r\,\sin b|/h}
\end{equation}
where $n_{0}$ is the local density in the mid-Plane.
If so, then the number of stars in a spherical volume of radius $D$ centered on the Sun assumed to lie in the midplane of the Galactic Plane, $N(D)$,  can be found from a Taylor series expansion as:
\begin{equation}
N(D)\;{\approx}\;\frac{4{\pi}\,n_{0}}{3}\,D^{3}\left(1\,-\frac{3}{8}\,\frac{D}{h}\,+\,\frac{1}{10}\frac{D^{2}}{h^{2}}\,-\,...\right)
\end{equation}
With $D$ = 80 pc and $h$ no larger than 150 pc \citep{Gilmore2000}, we expect to find no more than 91 DBs.  Because we identify 57 stars in Table 1, our sample is 
 at least    60\% complete.
   
   Most of the stars in Table 1 are listed in \citet{Bergeron2011} and \citet{Voss2007}.  When a star is considered in both samples, to be as consistent as possible, we adopt the stellar parameters in \citet{Bergeron2011}. For a few stars, only very limited data are available. 
   
  One of the most important DBs for our purposes is HS 2253+8023 because
   it has a large amount of external pollution.  The star's gravity is not well constrained in the detailed study of \citet{Klein2011}; we adopt their most probable value.  The distance
   to this star is derived from its radius, effective temperature and J-band 2MASS magnitude.  We adopt a similar approach to determine distances to
   the other stars whose distances are not provided by \citet{Bergeron2011}.     
   
   The average mass of the DBs in Table 1 is  0.67 M$_{\odot}$.  Using the initial mass final mass relation of \citet{Williams2009}, this implies a typical
   main-sequence progenitor mass of 2.5 M$_{\odot}$.

     \subsection{Upper Bounds to the Hydrogen Accretion Rates}
   We now consider a DB's hydrogen budget.  
   For the $i$'th star, we compute the upper bound to the hydrogen accretion rate averaged over the entire cooling age of the star,
   $\overline{{\dot M_{i}(H)}}$, simply as the total mass of accumulated hydrogen, $M_{i}(H)$, divided by the white dwarf's cooling age, 
 $t_{cool,i}$:
 \begin{equation}
{\overline{\dot M_{i}(H)}}\;=\;\frac{M_{i}(H)}{t_{cool,i}}
\end{equation}
As explained in \citet{Jura2011}, we expect the loss of accreted hydrogen by the action of a stellar wind to be negligible.  
 To determine $M_{i}(H)$, we have used  reported values of [H]/[He]  and  models  for the mass of the
   star's convective zone as a function of stellar gravity and effective temperature.   Because  \cite{Koester2009}  only
   presented results for one value of white dwarf gravity and hydrogen composition, D. Koester  kindly provided a grid of models for DB stars
   with a sufficient range of temperature, mass and [H]/[He] to enable us to interpolate for the mass of the  convective zone for each star individually in Table 1.  
  The estimates of $M_{i}(H)$ are strongly sensitive to
 the star's total mass. For example, for  DBs with an effective temperature of 13,000 K, the mass of the convective zone is  nearly a factor of 200 smaller in a star of 1.1 M$_{\odot}$ compared to a star of 0.6 M$_{\odot}$ \citep{Dufour2010}.   In most cases, we take [H]/[He] from the optical study of \citet{Bergeron2011}, but for a few stars hotter
 than 20,000 K, we use upper bounds from published ultraviolet observations which are appreciably more sensitive to the hydrogen abundance.  All our
 estimates for  $\overline{{\dot M_{i}(H)}}$ are upper bounds because the white dwarf could possess primordial hydrogen \citep{Bergeron2011}.

  As shown in Figure 1, for the seven stars in Table 1 for which both  \citet{Bergeron2011} and \citet{Voss2007} derived masses spectroscopically, instead of just adopting an assumed mean, the values of $\overline{{\dot M_{i}(H)}}$ are systematically lower from our use of the  \citet{Bergeron2011} study compared to those in \citet{Voss2007} by  an average factor of 4.
 This substantial discrepancy is the result of four factors all operating in the same direction. First, \citet{Bergeron2011} typically found larger stellar masses.  The typical 15\% increase  in derived mass means that the mass of the convective zone is  smaller by approximately a  factor of 2.  Second, stars with a larger mass take a longer time to cool
 to the observed effective temperature; this effect might contribute a factor of 1.2 to the estimate of $\overline{{\dot M_{i}(H)}}$.  Third,  because the star's gravity is estimated to be higher in the analysis of \citet{Bergeron2011}, the ratio of
 the hydrogen to helium may be found to be as much as a factor of 2 lower than derived by \citet{Voss2007}.  Finally, there is a different theoretical treatment of convection such
 that the values of the mass of the convective zone are typically a factor of 1.5 lower in  the models of \citet{Koester2009} compared to  those used by \citet{Voss2007}. 
 The spectroscopically-derived masses derived by \cite{Bergeron2011} usually agree very well with those derived from trigonometric parallaxes, and therefore it seems likely that the values of $\overline{{\dot M_{i}(H)}}$ derived from their analysis are realistic.  It must be recognized, however, that there are uncertainties.

 A few stars in Table 1 have remarkably low upper limits to $\overline{{\dot M_{i}(H)}}$.   These stars can be used to place a bound on the space density of interstellar comets \citep{Jura2011}.
  
 \subsection{Heavy Atom Accretion Rates}
 
 We wish to compare the hydrogen and heavy atom accretion rates.
For the $i$'th star and the $j$'th element, the time-averaged  mass accretion rate from the disk onto the star, $\overline{{\dot M_{i}(Z_{j})}}$, we take
\begin{equation}
\overline{\dot M_{i}(Z_{j})}\;=\;\;\frac{M_{i}(Z_{j})}{t_{i}(Z_{j})}
\end{equation}
where $M_{i}(Z_{j})$ is the inferred mass of element $Z_{j}$ in the star's mixing zone  where the  gravitational settling time is $t_{i}(Z_{j})$.   Using $Z_{tot}$ to denote the mass of all the heavy elements, we define for each star the time-averaged total heavy element accretion rate, $\overline{{\dot M_{i}(Z_{tot})}}$, as:
\begin{equation}
\overline{{\dot M_{i}(Z_{tot})}}\;=\;{\sum_j} {\overline{\dot M_{i}(Z_{j})}}
\end{equation}
 In many cases, only calcium is measured because this element is the most easily to detect in optical spectra; the total heavy element accretion rate is extrapolated as described 
 by \citet{Zuckerman2010}.    The values of $\overline{{\dot M_{i}(Z_{tot})}}$  are relatively insensitive to the white dwarf mass because, as illustrated in \citet{Klein2010}, the
 settling time scales approximately with the mass of the convective zone.

Three sources of hydrogen in DBs have been proposed: interstellar \citep{Voss2007}, circumstellar \citep{Jura2010} and primordial \citep{Bergeron2011}.    Because the heavy elements settle while the hydrogen does not, the absence of a correlation in Figure 2 is not a decisive test of any particular model for the source of the DBs' hydrogen.

 \subsection{Summed Accretion Rates}
   
Although there are nearby DBs that are missed, we assume that in identifying the stars, there is no bias related to their external-pollution.  If so,
then we can treat the set of stars as a well defined ensemble and  therefore compare the average hydrogen and heavy atom accretion with each other
in a meaningful manner.  We define ${\dot M_{Sum}}(H)$ as:
\begin{equation}
{\dot M_{Sum}(H)}\;=\;{\sum_i}{\overline{\dot M_{i}(H)}}
\end{equation}
For comparison,  the summed heavy atom accretion rate, ${\dot M_{Sum}(Z_{tot})}$, is:
\begin{equation}
{\dot M_{Sum}(Z_{tot})}\;=\;{\sum_i}\;\;{\overline{\dot M_{i}(Z_{tot})}}
\end{equation}

  With estimates in Table 1 of ${\overline{\dot M_{i}(Z_{tot})}}$ for 2/3 of the stars, and ignoring any contribution from the remaining 1/3 of the stars,  ${\dot M_{Sum}(Z_{tot})}$ is 1.6 ${\times}$ 10$^{10}$ g s$^{-1}$.  Nearly all  stars in Table 1 have been examined for hydrogen, and ${\dot M_{Sum}(H)}$ is ${\leq}$1.4 ${\times}$ 10$^{7}$ g s$^{-1}$, approximately a factor of 1000 smaller than the rate for heavy atoms.  Because some currently-unknown fraction  of this hydrogen
 is derived from the interstellar medium or is primordial, we conclude that in the entire ensemble  less than 1\% of the accreted mass is carried in water.  This bound is only
 strengthened if there is appreciable accretion of heavy atoms in those  stars which have not been examined with high spectral sensitivity.  Because we examined 57 stars, the upper bound to the average accretion rate for an individual star is 2.5 ${\times}$ 10$^{5}$ g s$^{-1}$.

  For both heavy elements and hydrogen, the asummed accretion is dominated by a few stars.  If we remove the top three accretors in
 each category, then ${\dot M_{Sum}(Z_{tot})}$ and ${\dot M_{Sum}(H)}$ are ${\leq}$2.6 ${\times}$ 10$^{9}$ g s$^{-1}$ and 4.4 ${\times}$ 10$^{6}$ g s$^{-1}$, respectively.  In this analysis, the ratio of the two accretion rates is approximately a factor of 600. Considered as an aggregate and interpreting the data with the best
 available parameters, we find that extrasolar asteroids accreted onto the warm DB stars are appreciably
 drier than CI chondrites.  
 
 For the four stars in Table 1 for which comprehensive abundance studies have been performed, we show in Figure 3 
   a plot of
the fraction of the total mass of each of the major individual constituents -- O, Mg, Si, Ca and Fe -- and the upper bound to the fraction of the toal mass that could have been   water.  In this Figure, we assume the steady state approximation given by Equation (4).    It can
be seen that there is a relatively small scatter in most of the mass fractions -- especially oxygen, consistent with our treatment of the extrasolar asteroids orbiting separate stars as one population.   We also see that the upper limit derived from the analysis of the aggregate population is significantly lower
than the  limit for any individual star.  The individual bounds plotted in Figure 3 are    derived  by using the lower of the two  values for the water fraction in the parent body.    The first value which is usually the stronger contraint assumes that all the oxygen in excess of that which could be bound into MgO, SiO$_{2}$ and Fe$_{2}$O$_{3}$ was contained in water. The second value assumes all the atmospheric hydrogen was bound into water. As explained in Section 3, the evidence that GD 61
has accreted ice-rich material is ambiguous, and we therefore plot the water fraction as an upper limit rather than as a measured quantity.  

We see in Figure 3 that the extrasolar asteroids are in aggregate drier even than CV chondrites -- the most refractory-rich of common meteorites \citep{Wasson1988}.
We also see that  commonly the extrasolar asteroids also are more rich in the refractory element Ca than the CV chondrites.  We can understand the dryness of extrasolar
asteroids as being a result of their having formed interior to the snow line.  We do not yet have a good model to explain the relative enhancement of Ca
that is measured in these systems.

  \section{MODELING INDIVIDUAL ACCRETION EVENTS}
  
  Having established statistically that ice is low in abundance within our sample of extrasolar asteroids, we now revisit model interpretations of elemental abundances in  individual  stars.
 Because different elements settle at different rates, there is 
not necessarily  a simple proportionality between the relative abundances in the atmosphere of the white dwarf and the fractional masses in the parent body.   \citet{Koester2009} described three regimes for an accretion event.    Initially, after an asteroid is tidally-disrupted into a disk,
there is a build-up of heavy elements in the outer convective zone of the star.  In this regime, the relative abundances in the stellar atmosphere directly scale
as the abundances in the parent body.  However, once the elapsed time since the onset of accretion becomes comparable to the gravitational settling
time,  the system enters a second phase.  In this situation, it has been  assumed that the gravitational settling rate balances the accretion rate and
an exact  steady state is established.   Finally, the disk material is completely accreted and the system enters a third phase when the  atmospheric pollution
decays with the elements with longer settling times lingering in the atmosphere and nominally appearing overabundant.   During the first and second
phases, the white dwarf possesses a cricumstellar disk.  In the third phase, the disk is dissipated.  If the disk is dusty, then we can detect an infrared
excess.  If, however, the disk is largely gaseous \citep{Jura2008} then it may not be observationally evident.

From the modeling point-of-view,
the second phase is the most uncertain because material is both being added and lost from the mixing zone.  In the first phase, material is only being
added and in the third phase, material is only being lost.  We therefore reconsider models for the second phase of an accretion event and reconsider the
assumption of an exact steady state.   The physics of the accretion from the disk onto the star is not fully understood; it might be variable.   \citet{Rafikov2011a} has argued that
Poynting-Robertson drag on the disk is important, but clearly this is only part of the story and the accretion rate may vary significantly \citep{Rafikov2011b, Belyaev2011} during a disk's evolution.  Observations of the time-variation and profiles of emission lines from the circumstellar matter \citep{Gaensicke2006, Gaensicke2007, Gaensicke2008, Melis2010} raise the possibility that there are significant changes in the accretion rate.   We now consider a quasi steady-state where there is a disk, but, nevertheless, the accretion rate  varies in time and 
the system is not in an exact steady state.

\citet{Dupuis1993} computed models for accretion of heavy atoms from the interstellar medium  onto DB white dwarfs.  Because they invoked
both high rates of accretion from clouds and low rates of accretion from the intercloud medium, their calculations explored the observational consequences of
time-varying accretion and  showed that the photospheric ratio of Mg to Ca could be dramatically enhanced because Mg gravitationally settles more
slowly.  While recent evidence demonstrates that  the heavy atoms in DBs mainly originate from asteroidal parent bodies, their finding that
the star's atmospheric abundances are sensitive to  the time-history of the entire system remains valid.
Our models extend the previous work of \citet{Dupuis1993}.

\subsection{Quasi-steady Accretion Model}

In the outer mixing zone of a white dwarf, the balance between accretion and settling is governed by the expression \citep{Koester2009}:
\begin{equation}
\frac{dM_{*}(Z_{j})}{dt}\;=\;-\frac{M_{*}(Z_{j})}{t_{j}}\,+\,{\dot M_{PB}}(Z_{j})
\end{equation}
where ${\dot M_{PB}}(Z_{j})$ denotes the accretion rate from  the circumstellar disk and is a measure
of the composition of the tidally-disrupted parent body\footnote{For notational convenience, in this Section, we denote the mass in the mixing zone in each star as $M_{*}(Z_{j})$ instead of $M_{i}(Z_{j})$ as in Section 2.}.  
Performing an integration on both sides and assuming that $M_{*}(Z_{j})$ = 0 at $t$ = 0, the solution to Equation (8) is:
\begin{equation}
M_{*}(Z_{j})\;=\;e^{-t/t_{j}}\,{\int}_{0}^{t}\,e^{t'/t_{j}}\,{\dot M_{PB}(Z_{j})}\,dt'
\end{equation}
Inverting the integral in Equation (9) is required to determine the parent body composition.
One  approach is to Fourier analyze  the accretion rate.  Because of the large unknowns,  for simplicity, we consider just the illustrative special case of a single Fourier term:
\begin{equation}
{\dot M_{PB}(Z_{j})}\;=\;{\overline{\dot M_{PB}(Z_{j})}}\,\left(1\,+\,r\,\sin{\omega}\,t'\right)
\end{equation}
Here $r$ and ${\omega}$ are free parameters that characterize the amplitude and frequency of the fluctuations in the accretion rate of the $j$'th element  whose time-average value is ${\overline{\dot M_{PB}(Z_{j})}}$. In addition to accretion rate variations intrinsic to the disk, there may be multiple disruptions of parent bodies
that also may lead to time variations of the accretion rate \citep{Jura2008}.  In all cases, we require $r$ $<$ 1.

For convenience, define the dimensionless term:
\begin{equation}
B_{j}\;=\;t_{j}\,{\omega}
\end{equation}
Then:
\begin{equation}
M_{*}(Z_{j})\;=\;{\overline{\dot M_{PB}(Z_{j})}}\;\;t_{j}\;\left(\left[1\,-\,e^{-t/t_{j}}\right]\,+\,\left[\frac{r}{1\,+\,B_{j}^{2}}\right]\left[\sin{\omega}\,t\;-\;B_{j}\cos{\omega}\,t\;+\;B_{j}\,e^{-t/t_{j}}\right]\right)
\end{equation}

It is instructive to consider some limiting case solutions to Equation (12).  When $t$ $<<$ $t_{j}$, then:
\begin{equation}
M_{*}(Z_{j})\;{\approx}\;{\overline{\dot M_{PB}(Z_{j})}}
\end{equation}
In this circumstance, the element's mass in the stellar atmosphere directly corresponds to the accretion rate; the initial build-up phase discussed by \citet{Koester2009}.  When
$t$ $>>$ $t_{j}$, then the solution to Equation (12) is:
\begin{equation}
M_{*}(Z_{j})\;{\approx}\;{\overline{\dot M_{PB}(Z_{j})}}\;\; t_{j}\;\left(1\,+\,\left[\frac{r}{1\,+\,B_{j}^{2}}\right]\left[\sin{\omega}\,t\;-\;B_{j}\cos{\omega}\,t\right]\right)
\end{equation}
In the exact steady state where the accretion rate is constant and $r$ = 0, then the abundance of an element in the parent body  are just given by the abundances
in the photosphere divided by the settling time and from Equation (14) we recover Equation (4).  However, if  $r$ $>$ 0,   the relationship between the photospheric concentration of an element and its relative abundance in the parent body is not the simple proportionality of Equation (4).

By examination of Equation (14), we can see that the quasi steady-state approximation usually agrees with the exact steady state model by better than a factor of 2.  That is if $B_{j}$ $>>$ 1, then clearly $M_{*}(Z_{j})$ is essentially constant and the limit of Equation (4) is reached.   In this case, variations in the accretion time are short compared to the settling time
and the pollution mass in the mixing zone is constant.  If $B_{j}$ $<<$ 1, then $M_{*}(Z_{j})$ might vary substantially, but  the variations in the accretion rate are so slow that the system's behavior is very similar to the
exact steady state.   The only situation when there is a significant difference between the quasi-steady state and the
exact steady state is when $B_{j}$ ${\approx}$ 1.  In this case, the settling and accretion times are
comparable and the interplay between the two factors can be complex, each element behaving differently according to its specific value of $B_{j}$.

\subsection{Was the Parent Body Accreting onto GD 61 Ice-Rich?}

We now use the model of quasi-steady accretion to address the question of whether the parent body accreted onto GD 61 was ice-rich as proposed
by \citet{Farihi2011a}.  
We see in Figure 3 that both G241-6 and GD 61 have relatively high fractions of oxygen in their contaminants, and therefore both systems are
candidates for the accreted parent body having had a substantial amount of water.  However, G241-6 has little photospheric hydrogen, and
therefore less than 10\% of the mass fraction of the pollution is water \citep{Klein2011}.  Because G241-6 does not have a circumstellar
dust disk \citep{Xu2011},  plausibly, it is in a late phase of an accretion event and the oxygen is especially abundant because it lingers
longer in the outer settling zone. In contrast, GD 61 has a substantial amount of hydrogen in its photosphere and a circumstellar dust
disk \citep{Farihi2011a}.  
  It is likely that there is ongoing accretion from the disk onto the star, and \citet{Farihi2011a} assume that
the GD 61 system is in an exact steady state.  If so, then by mass there is about 30\% more oxygen than can be locked into oxide minerals and this implies that the parent body contained a substantial amount of ice. This oxygen excess is sufficiently small that the analysis is sensitive  to whether the system is in an exact steady state or
only a quasi steady state.  

One difficulty with using the exact steady state  model to interpret  the data for GD 61 is that  iron displays a relatively low abundance.  
In an exact steady state model, as shown in Figure 3, it is inferred
to have an usually low fractional abundance compared to other polluted white dwarfs.   One possible way to explain simultaneously both the low iron and the high oxygen fractional abundances is that the system simply is in a late phase
where  only settling occurs \citep{Jura2010,Klein2011}.  However, this hypothesis  is inconsistent with the presence of a dust disk and therefore the likelihood that
accretion is ongoing.    \citet{Farihi2011a} suggested that the accretion onto GD 61 is iron-poor because the circumstellar disk is composed of the outer
portion of a planetesimal where most of the iron was concentrated into a central core.

GD 61 is deficient in carbon relative to the Sun by about a factor of 1000 \citep{Desharnais2008,Farihi2011a}.  According to \citet{Lee2010}, in planet-forming
disks, carbon should be treated as a volatile which therefore explains why this element is commonly observed to have a low abundance in
extrasolar asteroids \citep{Jura2006}.  If, in fact, the parent body accreted onto GD 61 was ice-rich, it would  be a puzzle to understand why it was simultaneously carbon-poor.

Here, we suggest that a quasi-steady state model may explain the data without any requirement that the accreted parent body contained ice.  Therefore, our input parameters   are chosen to find the largest effect of the quasi static model; they are not necessarily physically realistic.
Consequently, using Equation (14), we compute a model  with ${\omega}$ = 0.8 ${\times}$ 10$^{-5}$ yr$^{-1}$ and $r$ = 0.95.  Because the gravitational settling times typically are ${\sim}$10$^{5}$ yr \citep{Farihi2011a}, this  choice of ${\omega}$ means that the values of $B_{j}$ are nearly 1.   The relative mass fractions of O, Mg, Si and Fe are taken to equal 0.40, 0.18, 0.20 and 0.22; these relative abundances are 
close to  the values for bulk Earth \citep{Allegre2001} except for  iron which is about 50\%  too low.   However, because the  error in the photospheric abundance of Fe in GD 61 is 0.2 dex \citep{Farihi2011a}, the true fractional abundance of this element in the parent body is not required to be anomalously low compared to bulk Earth.

We display in Figure 4 the results of our model calculation for the fractional abundances in the photosphere at different times.  We see that there is a phase when the model matches the data.   It therefore seems that the argument that there must
be water in the parent body accreted onto GD 61 is provisional.

\section{DISCUSSION}

Hydrogen and oxygen are sufficiently cosmically abundant that in a disk where gas condenses into solids   at least
half of the total mass could be ice.  If, however, the temperature is high enough, only relatively refractory materials enter the solid phase and eventually
are incorporated into planetesimals.   Because the temperature in a disk decreases radially from the central star, it is likely that ice-rich objects form in cold, distant regions.   We find that water is less than
1\% of the mass of the ensemble of extrasolar asteroids accreted onto DB white dwarfs, and therefore these planetesimals  likely  formed interior to the snow line, the theoretical boundary between the ice-forming and the ice-free zones.

We have argued that oxygen is not sufficiently enhanced in GD 61's heavy element contamination to demonstrate convincingly that there was
water within the parent body.  In the future, a polluted white dwarf might be identified where the oxygen is sufficiently more abundant than the
other species that the argument for water could be more convincing.  In GD 40's photosphere, the number of O atoms is a factor of 2.3 times
greater than the sum of the number of Mg, Si and Fe atoms \citep{Klein2010}.  In this star, there is no evidence for any ice in the accreted parent body.
In GD 61's photosphere, the number of O atoms is a factor of 2.9 times greater than the sum of the number of Si, Mg and Fe atoms \citep{Farihi2011a}.
In this star, there may have been water in the parent body, but the case is uncertain. In the Sun, the number of O atoms is 4.9 times greater than the sum
of Mg, Si and Fe atoms \citep{Lodders2003}.  A planetesimal that formed in a very cold environment could have a correspondingly large
fraction of oxygen.  If this object is accreted onto a white dwarf,  its very high oxygen fraction could be measured and the case would be good
for an ice-rich parent body.  However, if the white dwarf is a DB, then the settling times are sufficiently long that unless there was also evidence for ongoing
accretion, it might be difficult to exclude the possibility that the enhancement of oxygen is simply the consequence of its lingering longer in the mixing zone.

We could hope to use warm  DA white dwarfs to study relative oxygen abundances because the settling times are shorter than a year and the system almost
certainly is in a steady state
 \citep{Koester2009}.    Unfortunately, the current situation is
murky.  GALEX J1931 is an exceptionally heavily polluted DA where all the major element contaminates are measured, and  the number of O atoms in the photosphere is somewhere between 1.0 and 1.8 times the sum of the Mg, Si and Fe
atoms \citep{Vennes2010, Vennes2011, Melis2011}.   The  two groups do not agree very well in their determination of the  iron abundance, perhaps because \citet{Vennes2010} only used one weak Fe line while \citet{Melis2011} used several iron lines.   More importantly for our purposes here, correcting for settling and consequent stratification through the atmosphere, the number of O atoms compared to the sum of Mg, Si and Fe atoms is either ${\sim}$11 \citep{Vennes2010} or ${\sim}$1 \citep{Melis2011}.  In the first case, likely the parent body  was ice-rich; in the second case  likely it was ice-poor. Because we have found that  ice-rich parent bodies are not ubiquitous; observations of additional DA stars may help resolve  this disagreement.   The analysis of  DB stars does not suffer this particular ambiguity.

In our toy model for external pollution, we have assumed that the ultimately-accreted hydrogen is chemically bound to oxygen in the form of water.  Another possibility is that this hydrogen could be chemically
bound to carbon in the form of hydrocarbons.  However,   carbon is highly volatile in planet-forming environments \citep{Lee2010} and deficient in those relatively few polluted white dwarfs where it has been studied \citep{Jura2006,Farihi2009}.    Therefore, H$_{2}$O is the most likely carrier of hydrogen in extrasolar minor planets.  This water could be contained in hydrated silicates such as serpentine rather
than pure ice \citep{Rivkin2002,Wasson2008}.

A large source of uncertainty in our estimates to the upper bounds to the hydrogen accretion rates  is the accuracy  of the adopted white
dwarf masses.  This difficulty should  be  eased with the {\it Gaia} mission which will determine trigonometric parallaxes of all nearby white
dwarfs, greatly extending the work of the {\it Hipparcos} satellite \citep{Vauclair1997}.

\section{CONCLUSIONS}

We argue that the aggregated external pollution from parent bodies accreted onto DB white dwarfs is less than 1\% by mass composed
of water.  Consequently, on average, the asteroids that  orbited main-sequence stars of typically 2.5 M$_{\odot}$ are  drier than CI chondrites by at least a factor of 20.  In contrast, we show that for an individual star such as GD 61, having a large fraction of oxygen in the polluted material does not
necessarily imply accretion from a water-rich parent body.  Despite the ambiguity associated with studying any single white dwarf, 
our analysis of the aggregate population of polluted DBs provides indirect observational support that snow lines in planet-forming disks are common.

We thank D. Koester for providing a grid of models for the masses of white dwarfs and for useful email exchanges.  We thank the anonymous referee
for thoughtful and constructive comments.  This work has been partly supported by the National Science Foundation.

\newpage
\begin{center}
Table 1 -- DBs Within 80 pc of the Sun
\begin{tabular}{lllllllllll}
\hline
\hline
\\
Star & Name&d & $b$ & T$_{eff}$ &  M$_{wd}$ &$\log$ $\overline{{\dot M(H)}}$ & $\log$ $\overline{{\dot M(Z_{tot})}}$ & Notes\\
(WD)& & (pc) & ($^{\circ}$) &(K) & (M$_{\odot}$) & (g s$^{-1}$) & (g s$^{-1}$) & \\
\hline
0002+729 & GD 408 & 32 & +11& 14,410 &0.75 &  ${\leq}$4.39 & 7.83  & (1), (2)\\
0017+136 & Feige 4 & 75 & -48 & 18,130 & 0.65 &${\leq}$5.24 &   &(1)  \\
0100-068 & G270-124 & 43 & -69 & 19,800 & 0.64 &${\leq}$3.63 &  7.70 & (1), (3)  \\
0125-236 & G274-39&61 &-81 & 16,610 &  0.75 &${\leq}$4.70 & 7.50 &(1), (2)\\
0138-559 & BPM 16571& 51 &-60 & 15,800 & && & (4) \\
\\
0300-013 & GD 40 &64&-49 & 15,300 &  0.67 & ${\leq}$4.51 &9.44 & (1), (5)\\
0308-565 & BPM 17088 & 50 &-52& 23,000 & 0.63 & $<$0.50 &$<$&(1), (6)\\
0414-0434 &HE & 61 &-36& 13,470 &  0.67 &${\leq}$5.14 &$<$ &(1), (7) \\
0418-539 & BPM 17731 &  76&-44 & 19,050 & 0.66 & $<$4.70 & $<$ &(1), (7) \\
0435+410 & GD 61 & 52&-04 & 16,810 & 0.70 & ${\leq}$5.91 & 8.81 & (1), (8)\\
\\
0437+138 & LP475-242& 45&-21 & 15,120 & 0.74 &${\leq}$5.63 &    &(1) \\
0503+147 & KUV & 31 & -15 &  15,610 & 0.64 & ${\leq}$5.20 & & (1) \\   
0517+771 & GD 435 &  69 &+22& 13,150 & 0.67 &${\leq}$4.80 & $<$5.92  &(1), (2) \\
0615-591 & NLTT 16355 &  37 &-27& 15,750 & 0.61 &$<$4.32 &$<$& (1), (7) \\
0716+404 & GD 85 & 59 &+22& 17,150 & 0.64 &$<$4.16 & $<$6.8&(1), (9)\\
\\
0840+262 & Ton 10 & 49&+35 & 17,770 & 0.78 &${\leq}$5.61 &     &(1) \\
0845-188 & NLTT 20260 & 75 &+15& 17,470 & 0.68 & $<$ 3.97 & $<$7.53 &(1), (2)\\
0948+013 & PG & 77&+40 & 16,810 & 0.65 &${\leq}$4.93 & $<$ &(1), (7) \\
1009+416 & KUV & 69&+55 & 16,480 & 1.00 & $<$4.00 & &(1)\\
1011+570 & GD 303 & 48&+49 & 17,350 & 0.67&$<$5.05 & 8.84&(1), (2) \\
\\
1046-017 & GD 124 & 70 &+48& 14,620 & 0.68 &$<$4.14 &  $<$6.20 &(1), (2)\\
1107+265 & GD 128 & 78&+67 & 15,060 &0.65 & ${\leq}$5.31 & 7.17 &(1), (2) \\
1129+373 & PG & 74&+70 & 13,030 & 0.68 &${\leq}$4.67 & $<$6.07 & (1), (2)\\
1333+487 & GD 325 & 35 &+67& 15,320 & 0.61 &$<$5.35 & &(1)\\
1336+123 & NLTT 34784 & 51 &+71& 15,950 & 0.60 &$<$4.41 & $<$&(1), (7) \\
\\
1352+004 & PG & 69 &+59& 13,980 & 0.62 & ${\leq}$5.60 & 7.37 & (1), (2)\\
1403-010 & G64-43 & 80&+56 & 15,420 & 0.65 &${\leq}$4.48 & $<$6.44 &(1), (2) \\
1411+218 & PG & 38&+71 & 14,910 & 0.62 &$<$5.34 & $<$6.26  &(1), (2)\\
1425+540 & G200-40 & 56&+58 & 14,490 & 0.56 & ${\leq}$6.86 & 7.73 &(1), (2) \\
1444-096 & PG & 52 &+44& 17,040 & 0.75 & ${\leq}$4.11 &$<$&(1), (7)  \\

\hline
\end{tabular}
\end{center}
 \newpage
 \begin{center}
 Table 1 -- Continued
 \\
 \begin{tabular}{lllllllllll}
\hline
\hline
\\
Star &Name & d&$b$ & T$_{eff}$ & M$_{wd}$ & $\log$ $\overline{{\dot M(H)}}$ & $\log$ $\overline{{\dot M(Z_{tot})}}$ & Notes\\
(WD) & &(pc)&($^{\circ}$) & (K) & M$_{\odot}$ & (g s$^{-1}$) & (g s$^{-1}$) & \\
\hline
1459+821 & G256-18 & 49&+34 & 15,850 & 0.65 & $<$5.20 & $<$6.36 &(1), (2)\\
1542+182 & GD 190&67&+49 & 22,630 & 0.63 & $<$0.58 &$<$&(1), (6)\\
1557+192 & KUV & 78 &+46& 19,570 & 0.68 & ${\leq}$4.48 &$<$& (1), (7)   \\
1610+239 & PG & 50&+45 & 13,360 & 0.69 & $<$5.13 &$<$6.11&(1), (2) \\
1644+198 & PG & 56&+36 & 15,190 & 0.66 & $<$5.17 & 6.76&(1), (2)\\
\\
1645+325 & GD 358 & 46 &+39& 24,940 & 0.57 & $<$1.14 & &(1), (10)\\
1703+319 & PG & 67 &+35& 14,430 & 0.88 & ${\leq}$4.37 &&(1) \\
1708-871 & BPM 921 & 58 &-26& 23,980 & 0.64 & $<$1.31 & &(1)\\
1709+230 &GD 205 & 65&+32 & 19,610 & 0.65 & ${\leq}$4.87 & 8.73&(1), (2) \\
1726-578 & BPM 24886 & 51&-13& 14,320 & 0.71 & ${\leq}$5.05 & &(1)  \\
\\
1822+410 & GD 378 & 45&+22 & 16,230 & 0.60 & ${\leq}$6.20 & 8.58 &(1), (2) \\
1919-362 & SCR J1920-3611 & 42&-21 & 27,800 &0.59 & & & (11) \\
1940+374 & NLTT 48137 & 47 &+07& 16,630 & 0.64 & $<$5.22 & $<$6.73 &(1), (2)\\
2034-532 & BPM 26944 & 35 &-37& 17,160 & 0.90 & $<$3.63 &&(1) \\
2129+000 & G26-10 & 49 &-35& 14,380 & 0.75 & $<$3.88 & $<$6.01&(1), (2)\\
\\
2130-047 & GD 233 & 50&-38 & 18,110 & 0.66 & ${\leq}$4.07 &  $<$7.40&(1), (2)\\
2144-079 &GJ837.1 &  49 &-42& 16,340 & 0.70 & $<$4.00 & 8.08&(1), (2)\\
2154-437 &BPM 44275 &  61&-52   &16,700 & 0.60 &${\leq}$5.68 &$<$ & (7) \\ 
2222+683 & G241-6 & 65&+10 & 15,230 & 0.71 &$<$4.97 & 9.30 &(1), (2)\\
2224-344 & LTT 9031 & 72  &-58  & 19,000 & & &  &(12)\\
\\
2229+139& PG & 76 & -37 & 14,940 & 0.70 & ${\leq}$5.75 & & (1) \\
2236+541 & KPD & 79 &-04& 15,470 & 0.78 & $<$4.76 & &(1)  \\
2253-062 & GD 243 & 63 &-55& 17,190 & 0.64 &${\leq}$5.90 & $<$ &(1), (7) \\
2253+8023& HS & 71 &+19 & 14,400 &0.84 &  ${\leq}$4.43 & 9.95&(13) \\
2310+175 & PG  & 64 &-39& 15,170 & 0.82 & $<$4.55 &&(1) \\
\\
2328+510 & GD 410 & 53&-10 & 14,460 & 0.63 & $<$5.30 &&(1) \\
2334-4127 & HE& 80 & -69 &18,250 & 0.61& ${\leq}$4.55 &$<$ & (7) \\
\hline
\end{tabular}
\end{center}
The entries in the column for $\overline{{\dot M(H)}}$ may be blank (no measurement reported in the literature), $<$ (no hydrogen detected, but an upper limit is reported) or ${\leq}$ (hydrogen detected, but the rate of accretion is an upper limit because there could be some primordial hydrogen).  We enter ``$<$" in the ${\log}$ $\overline{{\dot M(Z_{tot})}}$ column for those stars examined in the SPY survey  that do not display any calcium absorption but no quantitative upper limit is provided \citep{Voss2007}.
Notes: (1) \citet{Bergeron2011}; (2) \citet{Zuckerman2010}; 
(3) \cite{Desharnais2008};
(4) \citet{Sion1988};
(5) \citet{Klein2010}; 
(6) \citet{Petitclerc2005};
(7) \citet{Voss2007};
(8) \citet{Farihi2011a};
(9) \citet{Dupuis1993};
(10) \citet{Provencal2000};
(11) \citet{Subasavage2008};
(12) \citet{Castanheira2006}; 
(13) \citet{Klein2011};    

 \begin{figure}
 \plotone{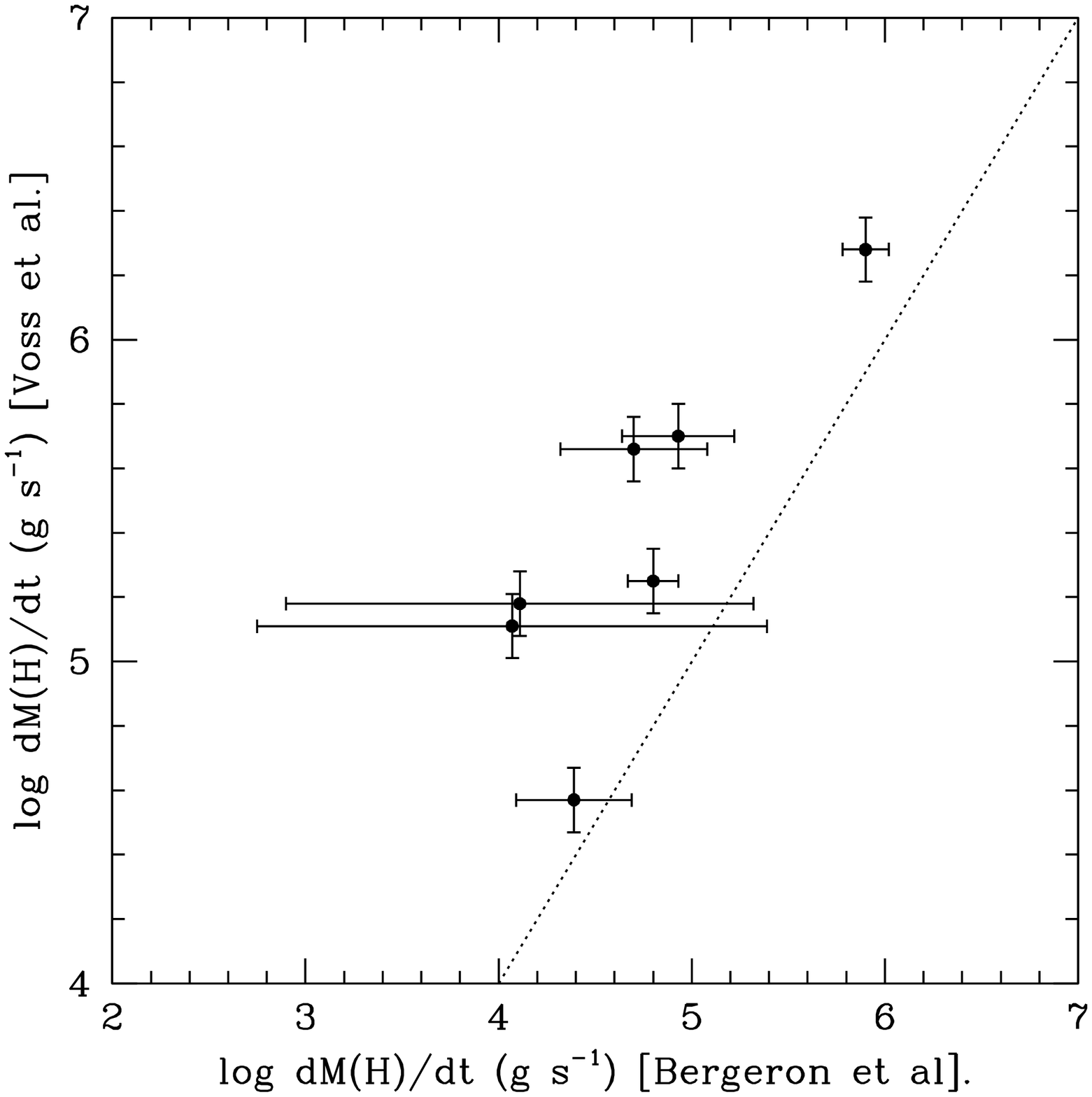}
 \caption{Comparison of time-averaged upper bounds to the hydrogen accretion rates, denoted ${\overline{\dot M(H)}}$ in the text, using the parameters in \citet{Voss2007} and \citet{Bergeron2011} for the seven stars in Table 1 for
 which both studies derived masses spectroscopically.  The dotted line displays the locus of points where the two  rates would agree.  As discussed in the text, the difference between the two studies is largely the result of \citet{Bergeron2011} deriving higher white dwarf masses. The errors are only those
 associated with the atmospheric abundance determinations; they do not reflect uncertainties in the stellar mass or effective temperature.  
 While \citet{Voss2007} found that the uncertainties in their derived hydrogen abundances might be as low as 6\%, our plotted error bars
 reflect our assumption that realistically, the errors  may be as large as 0.1 dex.   }
 \end{figure}
  \begin{figure}
 \plotone{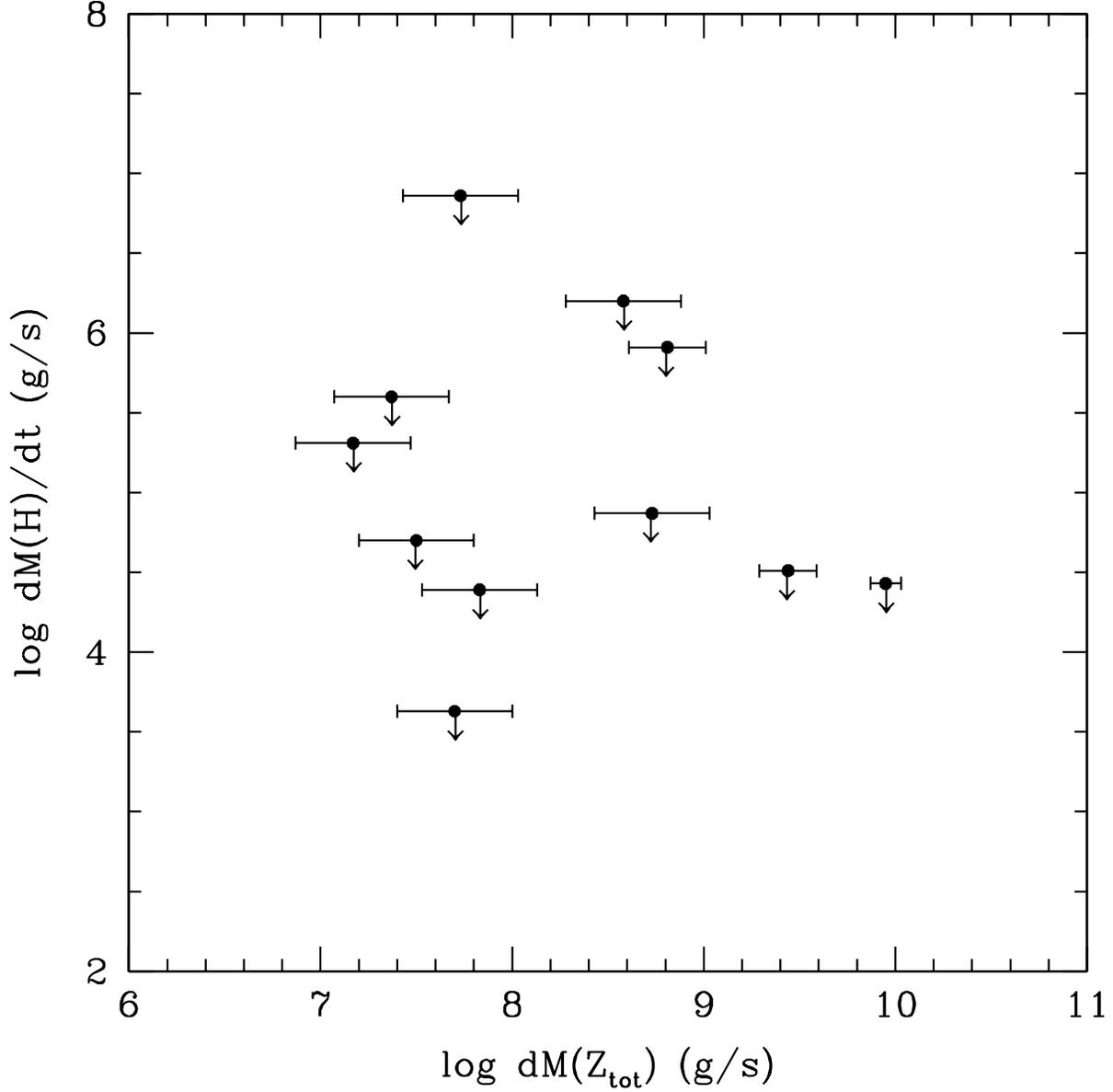}
 \caption{Comparison of  the upper bound to the time-averaged hydrogen accretion rates, denoted ${\overline{\dot M(H)}}$ in the text, with heavy atom accretion rates, denoted ${\overline{\dot M(Z_{tot})}}$ in the text,  for the stars in Table 1 where both quantities are determined.  For the stars where
 only the calcium abundance is reported, we assume an overall uncertainty of a factor of two in the total heavy-atom accretion rate.  For the
 stars with detailed abundance analysis, the plotted error bars are taken from the papers where the data are reported.  No correlation between hydrogen accretion and heavy atom accretion is evident.}
 \end{figure}

 \begin{figure}
 \plotone{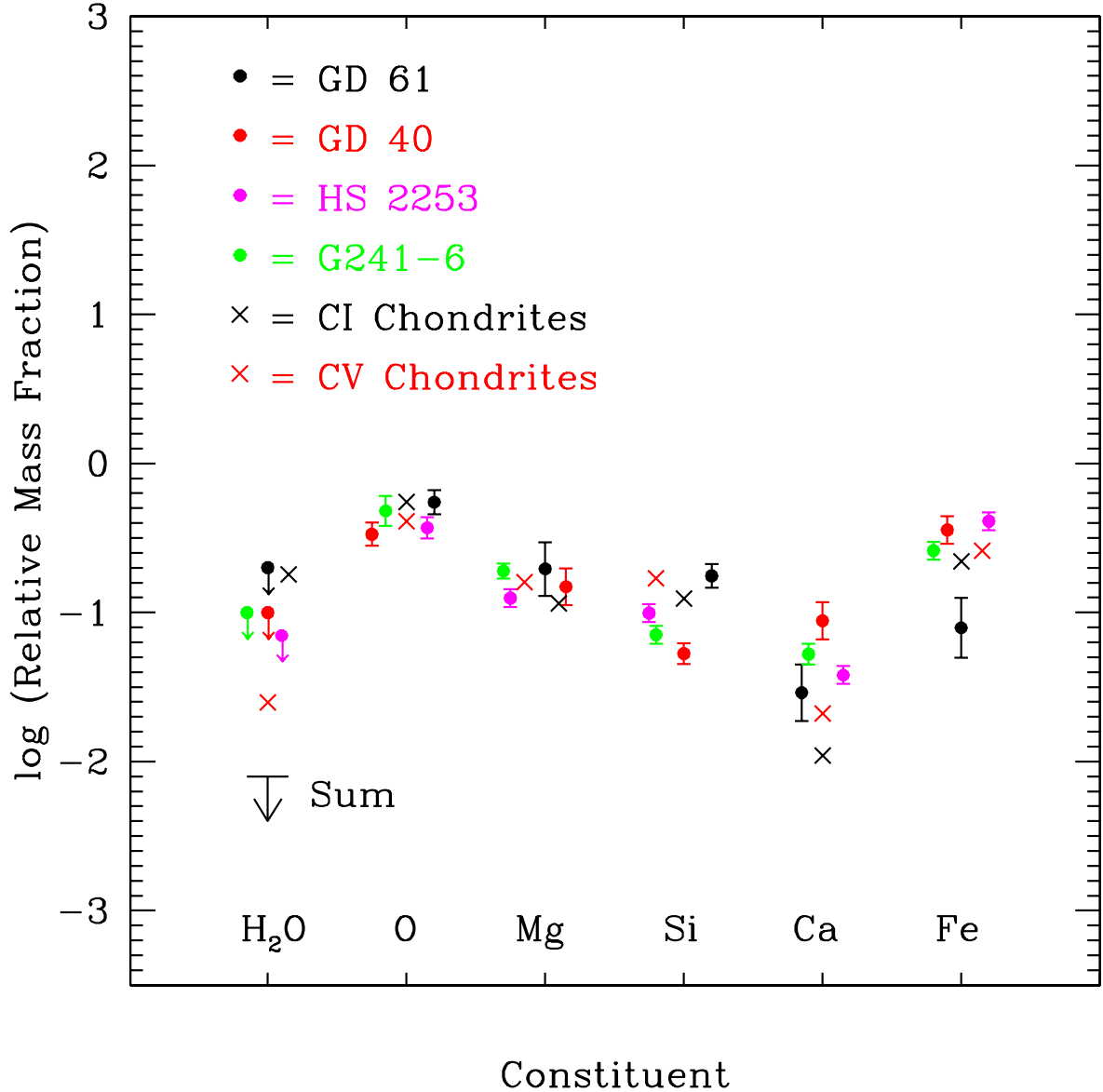}
 \caption{The relative fraction of the total mass of the accreted parent body carried by  individual constituents for the polluted white dwarfs in Table 1 where all  dominant elements -- O, Mg, Si, Ca and Fe -- have been measured.  The upper limit labeled with ``Sum" is the bound placed on the fractional aggregate water content of extrasolar asteroids in Section 2.  The  abundances  assigned to the parent bodies accreted onto  GD 61 \citep{Farihi2011a}, GD 40  \citep{Klein2010}, G241-6 \citep{Zuckerman2010} and HS 2253+8023 \citep{Klein2011} are all corrected from the atmospheric abundances by assuming a steady state and therefore using  Equation (4). The water fractions for the white dwarfs are all presented as upper limits; the case of GD 61 is
 discussed in detail in Section 3.2.  The water fractions for solar system CI and CV chondrites are shown for comparison; they are computed by assuming
 all the hydrogen in the meteorites is carried in water in the form of hydrated minerals.  }
 \end{figure}
 
 \begin{figure}
 \plotone{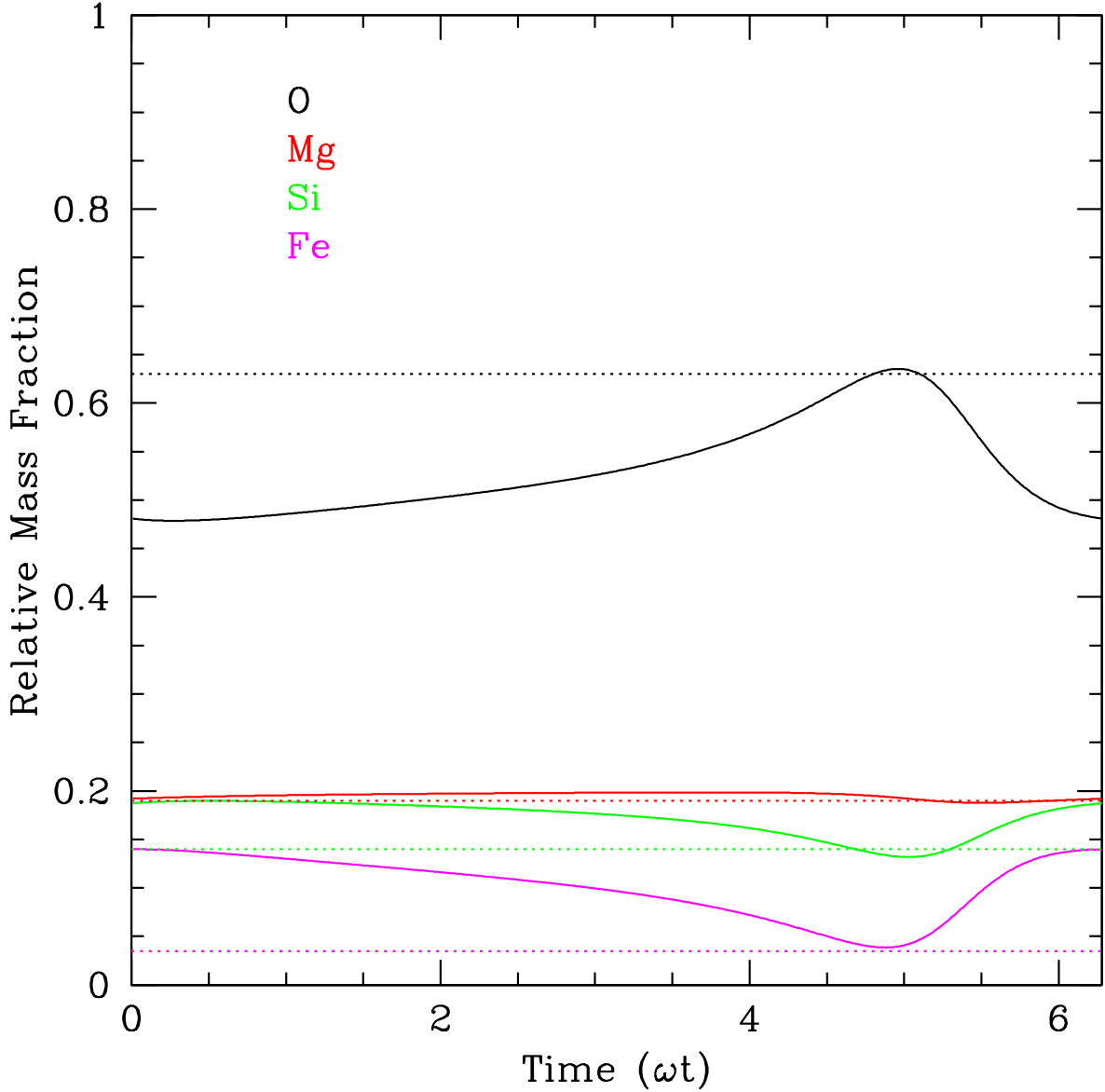}
 \caption{Predicted relative mass fractions of individual heavy elements from the accreted parent body within the photosphere of  GD 61 in the quasi-steady state model given by  Equation (13) with the parameters provided in the text.   The dotted horizontal lines display the values measured by \citet{Farihi2011a} and the solid curves the prediction as a function of time scaled to 
 dimensionless units.   Even though there is no ice in the model parent body, at ${\omega}t$ ${\approx}$ 4.9, the data are
 well matched by the model.  The computed fraction of the contamination that is oxygen is always greater than the assumed value in the parent body  of 0.4 because  oxygen settles
 relatively slowly compared to the heavier elements.}
\end{figure}
\end{document}